\documentclass[a4paper, journal]{IEEEtran} 
\IEEEoverridecommandlockouts
\usepackage[dvipsnames]{xcolor}
\usepackage{cite}
\usepackage{lipsum}
\usepackage{amsmath,amssymb,amsfonts, bm}
\usepackage{graphicx}
\usepackage{textcomp}
\usepackage{xcolor}
\usepackage{float}
\usepackage{cite}
\usepackage{placeins}
\usepackage{textcomp}
\usepackage{multirow}
\usepackage{xcolor}
\usepackage{wrapfig}
\usepackage{algpseudocode}
\usepackage{algorithm}
\usepackage{caption}
\usepackage{subcaption}
\usepackage{comment}
\usepackage[noabbrev]{cleveref}

\usepackage{tabularx, multirow,multicol,colortbl,array}
\usepackage{graphicx,color, array}
\usepackage[dvipsnames]{xcolor}
\usepackage{float}
\usepackage{comment,psfrag, url}
\usepackage{enumitem}
\usepackage{tikz,tikzscale,bigints,pgfplots}
\pgfplotsset{compat=1.18}
\usetikzlibrary{positioning, arrows.meta, colorbrewer, calc, fit, matrix, external}
\usepackage{multirow}

\usepackage{import}
\usepackage{pgfplots}
\usepgfplotslibrary{fillbetween}
\usetikzlibrary{calc}
\usetikzlibrary{spy}
\usetikzlibrary{matrix}
\usetikzlibrary{backgrounds}
\pgfplotsset{compat = newest}

\usetikzlibrary{shapes.misc, positioning, patterns}
\usetikzlibrary{shadows,arrows,shapes,calc,positioning,decorations.pathreplacing,fit}
\def\BibTeX{{\rm B\kern-.05em{\sc i\kern-.025em b}\kern-.08em
		T\kern-.1667em\lower.7ex\hbox{E}\kern-.125emX}}

\makeatletter
\def\endthebibliography{%
  \def\@noitemerr{\@latex@warning{Empty `thebibliography' environment}}%
  \endlist
}
\makeatother

\begin{document}
\pgfplotsset{
    standard/.style={
    axis line style = thick,
    grid = both,
    }
}

\pagestyle{empty}

\title{Phase-Aware Code-Aided EM Algorithm for Blind Channel Estimation in PSK-Modulated OFDM}

\author{
    \IEEEauthorblockN{ Chin-Hung Chen$^{\star}$}
    ,  \IEEEauthorblockN{Ivana Nikoloska$^{\star}$}
    , \IEEEauthorblockN{Wim van Houtum$^{\star\ddagger}$}%
    , \IEEEauthorblockN{Yan Wu$^{\ddagger}$}%
    , and \IEEEauthorblockN{Alex Alvarado$^{\star}$}%
    \\
    \IEEEauthorblockA{\textit{$^{\star}$Information and Communication Theory Lab, Eindhoven University of Technology, The Netherlands}}\\
    \IEEEauthorblockA{\textit{$^{\ddagger}$NXP Semiconductors, Eindhoven, The Netherlands}}\\
    \IEEEauthorblockA{c.h.chen@tue.nl}
}
\maketitle
\thispagestyle{empty}

\begin{abstract}
    This paper presents a fully blind phase-aware expectation–maximization (EM) algorithm for OFDM systems with the phase-shift keying (PSK) modulation. We address the well-known local maximum problem of the EM algorithm for blind channel estimation. This is primarily caused by the unknown phase ambiguity in the channel estimates, which conventional blind EM estimators cannot resolve. To overcome this limitation, we propose to exploit the extrinsic information from the decoder as model evidence metrics. A finite set of candidate models is generated based on the inherent symmetries of PSK modulation, and the decoder selects the most likely candidate model. Simulation results demonstrate that, when combined with a simple convolutional code, the phase-aware EM algorithm reliably resolves phase ambiguity during the initialization stage and reduces the local convergence rate from $80\%$ to nearly $0\%$ in frequency-selective channels with a constant phase ambiguity. The algorithm is invoked only once after the EM initialization stage, resulting in negligible additional complexity during subsequent turbo iterations.

\end{abstract}


\markboth{}%
{Shell \MakeLowercase{\textit{et al.}}: A Sample Article Using IEEEtran.cls for IEEE Journals}


\section{Introduction}\label{sec:intro}
    Robust channel estimation in frequency-selective fading environments is a fundamental requirement to achieve reliable wireless communication. For systems that do not use pilot symbols, blind estimation techniques are required, for which the expectation-maximization (EM) algorithm has become a widely used unsupervised method to obtain maximum-likelihood estimates, demonstrating strong performance in single-carrier systems \cite{Ghosh92, Kaleh94, Anton97}. To further enhance EM-based blind estimators, code-aided schemes that leverage extrinsic information from forward error correction (FEC) decoders have been proposed, demonstrating great performance improvements \cite{Lopes01, Garcia03, Zhao10, Niu05,Gunther05, CHC25}. However, EM-based algorithms are well known to be susceptible to initial conditions and often converge to a local maximum with poor initialization \cite{Yang12, CHC25_2}.

The primary causes of convergence to undesirable local solutions in EM algorithms for single-carrier systems are the inherent shift and phase ambiguities, which arise from the convolutional invariance of the inter-symbol interference (ISI) channel and the constellation symmetry of the transmitted symbols, respectively. \cite{Lopes01_2, Garcia03, CHC25_2}. In such cases, unreliable initial statistics can mislead the E-step of the EM algorithm, especially during the early iterations. To improve initialization, advanced methods such as semi-blind (data-aided) EM \cite{Kutz10, Nayebi18, Yang20} and deep learning-aided EM \cite{Schmid25} have been proposed to improve robustness against poor initialization. Despite their effectiveness, these approaches typically rely on pilot symbols or data-driven training, which is not compatible with a fully blind framework. Blind detection strategies based on differential PSK, which are not sensitive to a common phase ambiguity, have been proposed \cite{XMChen01, Serbetli14, WvH12}, but an effective blind phase detection method based on the EM algorithm for non-differential PSK remains an open challenge. 

In \cite{CHC25_2}, we proposed a joint shift and phase ambiguity detection algorithm for single-carrier systems by exploiting the decoder output. The key idea is to restructure the extrinsic likelihood function of the EM-based estimator–equalizer such that the decoder outputs a finite set of model evidences that serve as metrics for model selection. Ambiguities are then resolved by selecting the candidate with maximum model evidence. Although effective, this approach suffers from exponentially increasing computational complexity as the channel memory grows, leading to performance degradation as the number of candidate models required for joint ambiguity resolution increases.

Orthogonal frequency-division multiplexing (OFDM) provides a natural framework for combating frequency-selective fading by converting the linear convolution of the ISI channel into an element-wise multiplication in the frequency domain. This property simplifies equalization to a scalar operation per subcarrier, reducing the computational burden of EM since the hidden channel variables decouple across subcarriers. Consequently, EM-based estimation in OFDM systems has attracted significant attention \cite{Xie03, Wymeersch06, Ma04, Obradovic08, Ladaycia19}. However, phase ambiguity remains a major limiting factor for pilot-free blind EM algorithms in OFDM systems

This paper extends the blind ambiguity detection principle of \cite{CHC25_2} to PSK-modulated OFDM systems. Leveraging the structure of OFDM, the EM algorithm decomposes into independent subcarrier-wise operations, substantially reducing E-step complexity. Moreover, the shift ambiguity is effectively resolved as the ISI channel is reduced to a single tap per subcarrier. An EM update rule tailored for OFDM is formulated, and phase ambiguity resolution is incorporated through a model-evidence-driven detector supported by decoder extrinsic information. This integrated design provides a low-complexity approach to mitigating blind estimation phase ambiguity. Simulation results demonstrate that the refined initialization strategy substantially improves robustness and significantly reduces convergence to suboptimal local solutions, thereby enhancing the overall reliability of fully blind EM-based channel estimation in OFDM systems.

The remainder of this paper is organized as follows: Sec.~\ref{sec:sys} presents the transmission format. Sec.~\ref {sec:em} outlines the conventional EM algorithm design and the proposed phase-aware EM algorithm design. Sec.~\ref{sec:simu} presents the simulation results, and finally, Sec.~\ref{sec:conc} concludes this paper.

\section{OFDM Transmission Model}\label{sec:sys}
    \begin{figure*}
    \vspace{-10mm}
    \centering
    \resizebox{1\textwidth}{!}{\includegraphics{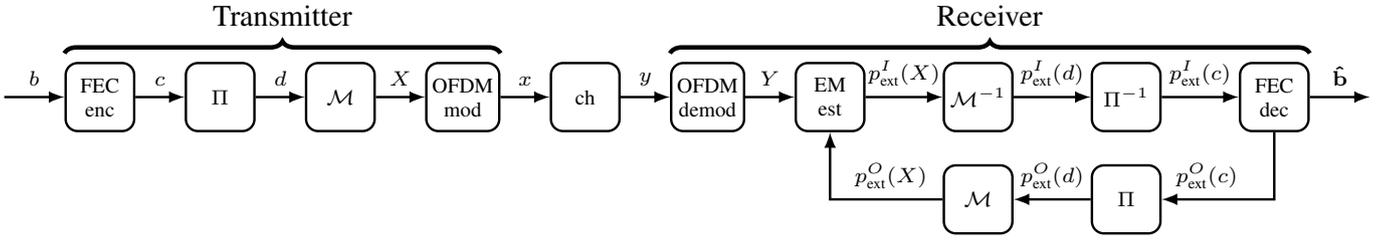}}
    \caption{System block diagram of a coded bit-interleaved PSK mapper with OFDM modulation on the transmitter side. The conventional receiver includes an OFDM demodulator, an EM estimator, and a turbo decoding module.}
    \label{fig:sys_block}
\end{figure*}

The notation convention used in this paper is defined as follows. Time-domain quantities are written in lowercase letters, while frequency-domain quantities are written in uppercase letters. Scalars denote individual entries (e.g., $x_k$ and $X_m$). Vectors are denoted by bold italic letters. For example, a time-domain vector from index $1$ to $K$ is written as $\boldsymbol{x}_1^K=[x_1, x_2, \ldots, x_K]^\mathsf{T}$, and a vector extracted from the $m$-th row and $n$-th column of a matrix is denoted by $\boldsymbol{X}_m$ and $\boldsymbol{X}^{(n)}$, respectively. Matrices are denoted by bold uppercase letters (e.g., $\mathbf{X}$), with $X_{m,n}$ representing the $(m,n)$-th entry. The transpose, Hermitian transpose, and complex conjugate are denoted by $(\cdot)^\mathsf{T}$, $(\cdot)^\mathsf{H}$, and $(\cdot)^\ast$, respectively.
We denote the circularly symmetric complex Gaussian distribution with mean $\mu$ and variance $\sigma^2$ as $\mathcal{CN}(\mu,\sigma^2)$, and its probability density function (pdf) evaluated at sample $x_t$ by $\mathcal{CN}(x_t;\mu,\sigma^2)$.

The transmission setup is shown in Fig.~\ref{fig:sys_block}. The information bit sequence $\boldsymbol{b}_1^{KR}$ is first processed by an FEC encoder that produces the encoded bit sequence $\boldsymbol{c}_1^{K}$.\footnote{A simple convolutional code is used in this work for illustrative purposes; however, the proposed methodology can be generalized to other FEC schemes.} Here, $R$ represents the coding rate and $K$ is the total length of the codeword per transmission frame. A bit-level interleaver ($\Pi$) is used to permute the FEC encoder output, where the interleaved coded bit sequence is represented as $\boldsymbol{d}_1^{K}$. Then, a PSK symbol mapper maps the binary coded bits to the sequence $\boldsymbol{X}_1^T$ where 
\begin{align} \label{eq:X}
    X_t \in \mathcal{\mathcal{X}}=\{s_i=e^{j2\pi i /C} \mid i=0,1,\dots,C-1 \}  
\end{align}
with $C=|\mathcal{X}|$ represents the PSK symbol cardinality.

In an OFDM system, the symbol information is encoded in the subcarriers. Let $M$ be the number of subcarriers and $N$ the number of OFDM symbols. The 1D PSK symbol vector $\boldsymbol{X}_1^T$ is reshaped to a 2D matrix $\mathbf{X}\in\mathbb{C}^{M\times N}$ with row $m$ represent the $m$-th subcarrier and column $n$ denotes the $n$-th OFDM symbol. The OFDM modulator then performs the inverse discrete Fourier transform (IDFT) to generate the time domain OFDM symbols via $\mathbf{x} = \mathbf{F}^H \mathbf{X}$, where $\mathbf{F}^H$ is the $M$-point IDFT matrix. We then append $N_{cp}$ cyclic prefix samples in front of each OFDM symbol to prevent ISI. Finally, the appended 2D time domain matrix is converted back to a 1D vector $\boldsymbol{x}_1^{M(N_{cp}+N)}$ for transmission.

We consider a frequency-selective channel where the input-output relationship can be expressed as
\begin{align}
    y_t = \sum_{l=0}^{L-1} h_{l} x_{t-l}  + w_t, \nonumber
\end{align}
where $h_l$ is the time-domain channel coefficient. We assume the noise realizations $w_t \sim \mathcal{CN}(0,\sigma_w^2)$ to follow a complex Gaussian distribution with zero mean and variance $\sigma_w^2$. During OFDM demodulation, the cyclic prefix of each OFDM symbol is first removed. The 1D received signal is then reshaped into a 2D matrix $\mathbf{y} \in \mathbb{C}^{M\times N} $, which is followed by the DFT
\begin{align}\label{eq:ch_f}
\mathbf{Y} = \mathbf{F} \mathbf{y}=\mathbf{H} \odot \mathbf{X} + \mathbf{W},
\end{align}
where $\odot$ denotes element-wise multiplication, $\mathbf{W}$ is the transformed AWGN matrix with entries $W_{m,n}\sim\mathcal{CN}(0,\sigma_w^2)$, and $\mathbf{H}$ denotes the channel frequency response matrix, which can be decomposed as
\begin{align}\label{eq:ch}
    H_{m,n} = |H_{m,n}| \, e^{j\phi_{m,n}},
\end{align}
where $|H_{m,n}|$ and $\phi_{m,n}$ denote the channel gain and phase respectively. Under the assumption of a frame-wise time-invariant channel, this simplifies to $H_m = |H_m| e^{j\phi_m}$.

\section{EM-based Estimation in OFDM}\label{sec:em}
    
\subsection{Conventional EM}
The conventional EM algorithm is well established in the literature \cite[Ch.~9]{Bishop}, \cite{HMM}, and its extension to OFDM-based channel estimation is presented clearly in \cite{Ma04}. For completeness, we briefly summarize the main steps of the EM algorithm for the OFDM system below. 

In the context of OFDM, the unknown parameters are the per-subcarrier channel coefficients $H_m$, and the hidden variables correspond to the transmitted symbols in $X_{m,n}$ with realization $s_i\in\mathcal{X}$ defined in \eqref{eq:X}. In OFDM systems, the estimation problem decomposes into independent subcarrier-wise updates where $H_m$ is estimated through $\boldsymbol{Y}_m=(Y_{m,1},Y_{m,2},\cdots,Y_{m,N})$. This significantly reduces computational complexity compared to single-carrier systems with the same number of observations $MN$. The EM algorithm iteratively alternates between two steps: (i) the expectation (E-step), which computes the posterior distribution of the hidden variables given the current parameter estimates $\hat{H}^{(p)}_m$ as $p(\boldsymbol{X}_m \mid \boldsymbol{Y}_m, \hat{H}^{(p)}_m)$ with the superscript $(p)$ denoting the $p$-th EM iteration, and (ii) the maximization (M-step), which updates the parameters by maximizing the expected complete-data log-likelihood
\begin{align}\label{eq:Q}
Q(H_m \mid \hat{H}_m^{(p)}) \triangleq \mathbb{E}_{\boldsymbol{X}_m\mid \boldsymbol{Y}_m,\,\hat{H}_m^{(p)}} \log p(\boldsymbol{Y}_m, \boldsymbol{X}_m\mid H_m).  
\end{align}

\subsubsection{E-step: posterior distribution inference}
Given the channel model in \eqref{eq:ch_f}, the likelihood function per subcarrier can be written as
\begin{align}
    p(\boldsymbol{Y}_m \mid \boldsymbol{X}_m, \hat{H}^{(p)}_m) 
    &=\prod_{n=1}^N \mathcal{CN}(Y_{m,n};\,\hat{H}_m^{(p)} X_{m,n}, \sigma_w^2), \nonumber 
\end{align}
The posterior probability that the transmitted symbol is equal to $ s_i \in \mathcal{X}$ given the current channel estimate $\hat{H}_m^{(p)}$ is
\begin{align} \label{eq:weight}
    &p_\text{pos}^{(p)}(X_{m,n}=s_i) \triangleq p(X_{m,n}=s_i \mid Y_{m,n}, \hat{H}^{(p)}_m) \nonumber \\ 
    &= \frac{p_{\text{pri}}(X_{m,n}=s_i)\cdot\mathcal{CN}(Y_{m,n};\, \hat{H}_m^{(p)} s_i, \sigma_w^2)}{\sum_{j=1}^C p_{\text{pri}}(X_{m,n}=s_j)\cdot \mathcal{CN}(Y_{m,n};\, \hat{H}_m^{(p)} s_j,\sigma_w^2 )},
\end{align}
where we define $p_{\text{pri}}(X_{m,n})$ as the prior information on $X_{m,n}$ and $p_\text{pos}(X_{m,n})$ as the posterior probabilities, which also serve as soft weights in the subsequent M-step.

\subsubsection{M-step: maximum likelihood parameter estimation}
Given the posterior symbol probabilities calculated at the $(p)$-th iteration in \eqref{eq:weight}, the expected log-likelihood \eqref{eq:Q} reduces to a weighted least-squares problem. The update for the channel coefficient at subcarrier $m$ is obtained as \cite{Ma04}
\begin{equation}
\tilde{H}^{(p+1)}_m 
= \frac{\sum_{n=1}^N \sum_{i=1}^C p_\text{pos}^{(p)}(X_{m,n}=s_i) \cdot (Y_{m,n} s_i^{*}) }
{ \sum_{n=1}^N \sum_{i=1}^C p_\text{pos}^{(p)}(X_{m,n}=s_i) \cdot( |s_i|^2) }.
\end{equation}
This corresponds to the weighted least-squares estimate of $H_m$ using the soft symbol assignments. 

Collecting all subcarrier-wise estimates, the channel frequency response vector ${\tilde{\boldsymbol{H}}}^{(p+1)}$ can be transformed into the time-domain impulse response by an IDFT 
\begin{align} \label{eq:hem1}
\tilde{\boldsymbol{h}}^{(p+1)} = \mathbf{F}^{H} {\tilde{\boldsymbol{H}}}^{(p+1)}
\end{align}
. After truncating to the assumed channel length $L$ with
\[
\boldsymbol{\hat{h}}^{(p+1)}=
\big[\tilde h^{(p+1)}_1,\,\tilde h^{(p+1)}_2,\,\ldots,\,\tilde h^{(p+1)}_L,\,
\underbrace{0,\,\ldots,\,0}_{M-L}\big]^\top,
\]
the channel coefficients $\hat{\boldsymbol{h}}^{(p+1)}$ are then mapped back to the frequency domain via DFT
\begin{align} \label{eq:Hem}
    \boldsymbol{\hat{H}}^{(p+1)} = \mathbf{F} \boldsymbol{\hat{h}}^{(p+1)} .
\end{align}
This refinement is presented in \cite{Ma04} which imposes a finite impulse response of length $L$,  thereby suppressing estimation noise. The EM iterations are repeated until convergence or until the maximum number of iterations is reached.
\subsection{Code-Aided EM}\label{sec:code_em}
The code-aided EM algorithm makes use of the extrinsic information from the decoder further to enhance the estimation performance for each turbo iteration \cite{Lopes01, Garcia03, Zhao10, Niu05,Gunther05, CHC25}. The extrinsic information refers to the information provided by one decoder (equalizer) about a specific bit (symbol), excluding the information used to derive it directly. Based on the soft posterior probability $p_\text{pos}(X_{m,n})$ obtained from \eqref{eq:weight} with the estimates ${\hat{H}}_m$, we first compute the extrinsic information as  
\begin{align} \label{eq:extx}
     p^I_{\text{ext}}(X_{m,n}) \triangleq {p_\text{pos}(X_{m,n})}/{p_{\text{pri}}(X_{m,n})},
\end{align}
where we set $p_{\text{pri}}(X_{m,n}) = 1 / C$ with equally prior in the first turbo iteration. Note that in this study, we use $p^I_{\text{ext}}(\cdot)$ and $p^O_{\text{ext}}(\cdot)$ to denote the extrinsic information generated from the inner code (i.e., the PSK demodulator) and outer code (i.e., the FEC decoder), respectively.
After symbol-to-bit demapping $p^I_{\text{ext}}(d_k) = \mathcal{M}^{-1}\big[p^I_{\text{ext}}(X_{m,n})\big]$ and de-interleaving $p^I_{\text{ext}}(c_{k}) = \Pi^{-1} \big[ p^I_{\text{ext}}(d_k) \big]$, The extrinsic information on the coded bits is then fed to the MAP decoder to generate the posterior probability on the coded bit as 
\begin{align}
    p_{\text{pos}}(c_k) \triangleq p(c_k \mid \boldsymbol{Y}_m) = \mathcal{F}\big[p^I_{\text{ext}}(\boldsymbol{c}_{1}^K),\, p_{\text{pri}}(\boldsymbol{b}_1^{KR})\big], \label{eq:jointc}
\end{align}
where $\mathcal{F}$ represents the MAP decoder function that takes likelihood  $p^I_{\text{ext}}(\boldsymbol{c}_{1}^K)$ and prior $p_{\text{pri}}(\boldsymbol{b}_1^{KR})$ input sequences. The detailed derivation of $\mathcal{F}$ for a convolutional decoder is provided in \cite{CHC25,CHC25_2}.
This extrinsic information from the decoder can then be obtained by dividing $p_{\text{pos}}(c_k)$ by the incoming extrinsic information from the symbol demodulator as
\[
    p^O_{\text{ext}}(c_k) = p_{\text{pos}}(c_k) / p^I_{\text{ext}}(c_{k}).
\] 
After bit interleaving $p^O_{\text{ext}}(d_k) = \Pi \big[ p^O_{\text{ext}}(c_k) \big]$, the extrinsic prior information is derived via a bit-to-symbol mapping
\begin{align}\label{eq:ext_x}
    p^O_{\text{ext}}(X_{m,n}) = \mathcal{M}\big[p^O_{\text{ext}}(d_k)\big].
\end{align}
Finally, \eqref{eq:ext_x} is fed back to the EM estimator to replace $p_{\text{pri}}(X_{m,n})$ on the right-hand side of \eqref{eq:weight} in the next turbo iteration. Note that we refer to the turbo iteration as the entire joint estimation and turbo equalization loop, while the EM iteration refers to the optimization within the EM estimator.

\begin{figure}
    \centering
    \resizebox{1\columnwidth}{!}{\includegraphics{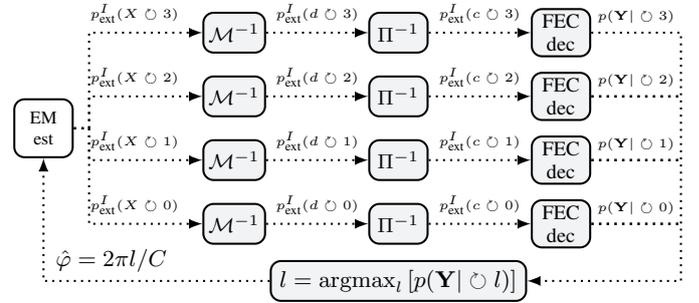}}
    \caption{Block diagram of the phase-aware code-aided EM algorithm design for a QPSK-modulated OFDM system.}
    \label{fig:sys_block_phs}
\end{figure}

\begin{figure*}[t]
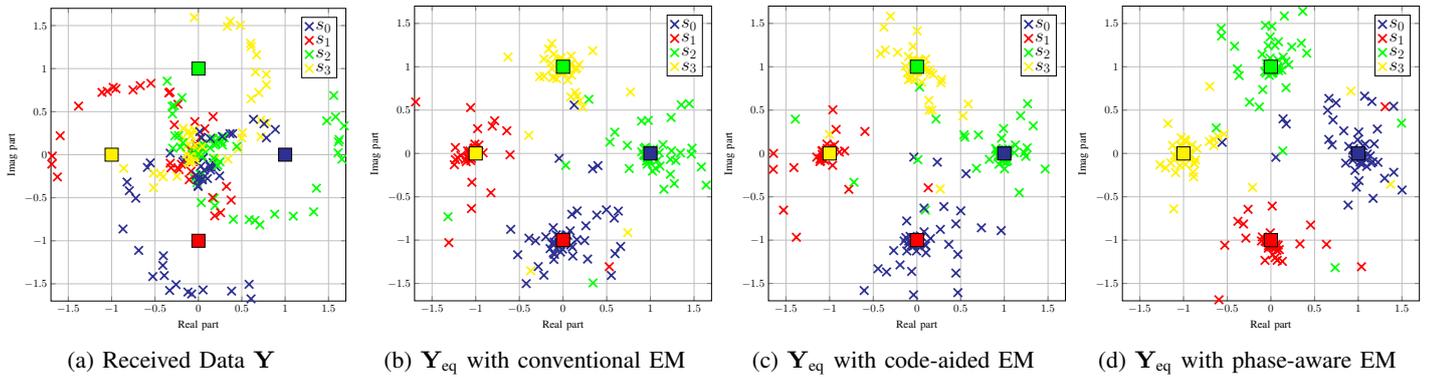

    \subfloat[Received Data $\mathbf{Y}$\label{fig:eq_Y}]{
        \centering
        \includegraphics[width=.25\textwidth]{fig/Y.tikz} 
    }
    \subfloat[$\mathbf{Y}_{\text{eq}}$ with conventional EM\label{fig:eq_std}] {
        \centering
        \includegraphics[width=.25\textwidth]{fig/EQ_std.tikz}  
    }
    \subfloat[$\mathbf{Y}_{\text{eq}}$ with code-aided EM\label{fig:eq_code}]{%
        \centering
        \includegraphics[width=.25\textwidth]{fig/EQ_code.tikz} 
    }
    \subfloat[$\mathbf{Y}_{\text{eq}}$ with phase-aware EM\label{fig:eq_phs}] {%
        \centering
        \includegraphics[width=.25\textwidth]{fig/EQ_phs.tikz}  
    }
    \vspace{0.5mm}
    \caption{One realization of constellation diagrams (crosses) of (a) the received symbols $\mathbf{Y}$ after the OFDM demodulation, and the equalizer outputs $\mathbf{Y}_{\text{eq}}$ using channel estimates obtained from (b) the conventional EM, (c) the code-aided EM, and (d) the phase-aware EM algorithms at $\text{SNR}=20$~dB. The four square markers represent the ideal symbol mapping $s_i$.}
\end{figure*}
\subsection{Phase-Aware Code-Aided EM}
In a symmetric constellation like PSK in \eqref{eq:X}, $C$ possible phase shifts result in the same constellation shape, leading to phase ambiguity. Consequently, EM-based blind estimators cannot effectively address these ambiguities. In \cite{CHC25_2}, an ambiguity detection algorithm was proposed to address both shift and phase ambiguities in single-carrier systems. In this work, we extend and develop this technique to OFDM, where the absence of ISI inherently resolves the shift ambiguity. We therefore focus on the phase-aware detection adapted to the OFDM system, as summarized below.

In Fig.~\ref{fig:sys_block_phs}, we show the schematic example of the phase-aware code-aided EM algorithm for a QPSK-modulated OFDM systems. The key to the algorithm is to utilize the decoder's sensitivity to the symbol-to-bit mapping to detect phase ambiguity effectively without relying on the pilot data. The algorithm first permutes the order of \eqref{eq:extx} and generates $C$ possible extrinsic symbol information as 
\begin{align}\label{eq:phs_ext}
    p^I_{\text{ext}}(X_{m,n}=s_i\circlearrowright l), \quad l=\{0,1,\ldots,C-1\},
\end{align}
where we use $\circlearrowright$ to denote the circular shift operation.
This information is then passed onto $C$ parallel symbol demodulators, deinterleavers ($\Pi^{-1}$), and convolutional decoders as described in Sec.~\ref{sec:code_em} to derive $p_{\text{pos}}(c_k \mid \circlearrowright l)$ as in \eqref{eq:jointc}. Under the assumption of a frame-wise constant phase ambiguity, the model evidence of the entire observation matrix $\mathbf{Y}$ is then computed for each of the parallel processing modules as
\begin{align}\label{eq:phs_evi}
    p(\mathbf{Y}|\circlearrowright l) \propto \prod_{m=1}^M\sum_{c_k}p_{\text{pos}}(c_k |\circlearrowright l),
\end{align}
where we determine the phase ambiguity via
\begin{align} \label{eq:phs_corr}
    \hat{\varphi} = {2\pi l/C}, \quad l=\operatorname*{argmax}\limits_{l\in \{0,1,\dots,C-1\}}{\big[p(\mathbf{Y}|\circlearrowright l)\big]}. 
\end{align}
Finally, we multiply the channel response estimated through \eqref{eq:Hem} by the phase detected through \eqref{eq:phs_corr} as $\boldsymbol{\hat{H}}e^{-j\hat{\varphi}}$ to mitigate the phase ambiguity.

\subsection{Computational Complexity}
We now analyze the computational complexity of the proposed phase-aware code-aided EM receiver for OFDM systems. We concentrate on the primary complexity of the EM algorithm and the MAP decoder, using the big-O notation $\mathcal{O}(\cdot)$ to examine how computation scales with the problem size. Let $MN$ be the total number of observations per OFDM frame and $N_{\text{EM}}$ for the total number of EM iterations. In each EM iteration, the E-step requires evaluating $C$ likelihoods per symbol, leading to $\mathcal{O}(MNC)$ while the M-step involves weighted accumulations of the same order. In addition, the channel refinement \eqref{eq:hem1}--\eqref{eq:Hem} is enforced via IDFT/DFT projection, with complexity $\mathcal{O}(M\log M)$ via fast Fourier transform (FFT). Hence, the EM equalizer cost 
\begin{align} \label{eq:O_em}
    \mathcal{O}\big(N_{\text{EM}}(MNC + M\log M)\big).    
\end{align}
When a convolutional code is used as the FEC scheme, the decoder processes $MN\log_2 C$ coded bits with $2^{L_c-1}$ states, yielding $\mathcal{O}\!\big(MN\log_2 C \cdot 2^{L_c-1}\big)$.
Here, $L_c$ denotes the constraint length of the convolutional code. Therefore, the overall per-turbo iteration complexity is
\[
\mathcal{O}\!\Big(
N_{\text{EM}}(MNC + M\log M) + MN\log_2 C \cdot 2^{L_c-1}
\Big).
\]

The proposed initialization step requires evaluating additional $C-1$ phase candidates using the decoder’s model evidence compared to the conventional code-aided EM. This introduces a one-time overhead of
\[
\mathcal{O}\!\Big(
(C-1) \big[MN\log_2 C \cdot 2^{L_c-1}\big]
\Big).
\]
This cost occurs only once at the initialization stage and does not repeat in the subsequent turbo iterations.

In contrast to the physics-aware EM system for single-carrier transmission in~\cite{CHC25_2}, where the EM equalizer suffers from exponential trellis complexity $\mathcal{O}(N_{\text{EM}} M N C^{L})$ due to ISI, the OFDM structure adopted here removes this exponential dependence and ensures that the complexity grows only linearly with the modulation order~$C$, as shown in~\eqref{eq:O_em}, while remaining independent of the channel length~$L$. Moreover, the shift ambiguity detection with complexity $\mathcal{O}\!\Big( L\big[MN\log_2 C \cdot 2^{L_c-1}\big]\Big)$ presented in~\cite{CHC25_2} is no longer required in the OFDM system. Consequently, the proposed phase-aware EM scheme for OFDM achieves a substantial complexity reduction.
\section{Simulation Result}\label{sec:simu}

\begin{figure*}[t]
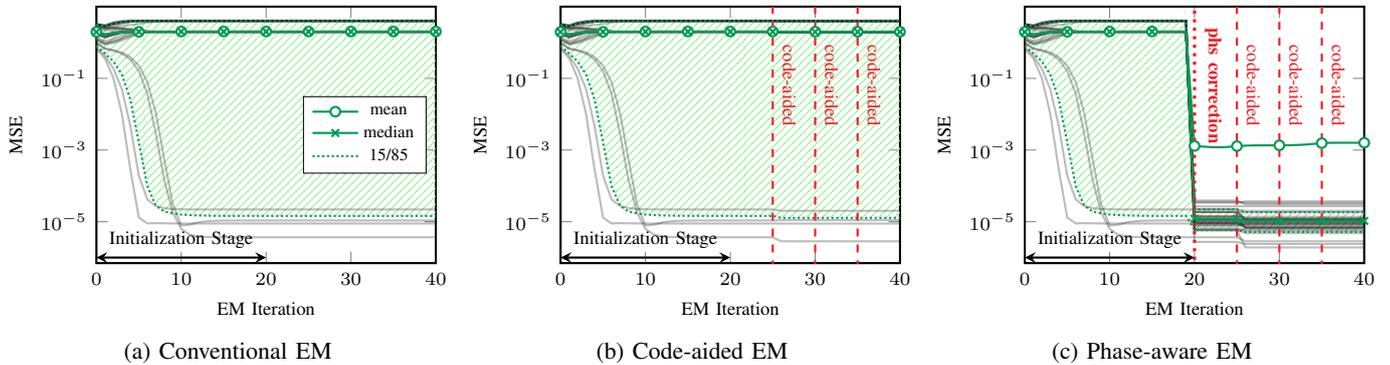

    \vspace{-5pt}
    \subfloat[Conventional EM\label{fig:e05_std}]{%
        \centering
        \includegraphics[width=.33\textwidth]{fig/mse_std.tikz} 
    }
    \subfloat[Code-aided EM\label{fig:e05_phs}]{%
        \centering
        \includegraphics[width=.33\textwidth]{fig/mse_code.tikz} 
    }
    \subfloat[Phase-aware EM \label{fig:e05_joi}]{%
        \centering
        \includegraphics[width=.33\textwidth]{fig/mse_phs.tikz}  
    }
    \vspace{0.5mm}
    \caption{Mean square error of the estimated channel frequency response from (a) conventional EM, (b) code-aided EM, and (c) the phase-aware EM. Gray lines indicate individual realizations (30 out of 5000) for an SNR of $20$~dB. The first 20 EM iterations are the initialization stage, where phase correction (red dotted line in (c)) is only performed at the 20th EM iteration. After initialization, extrinsic information from the decoder is incorporated every 5 EM iterations (red dashed lines in (b) and (c)). The mean and median of the 5000 independent simulations are represented by solid green circles and crosses, respectively, while the shaded areas indicate the 15th to 85th percentiles.}
\end{figure*}

\begin{figure}[t]
    \centering
    \includegraphics[width=1\columnwidth, height=.65\columnwidth]{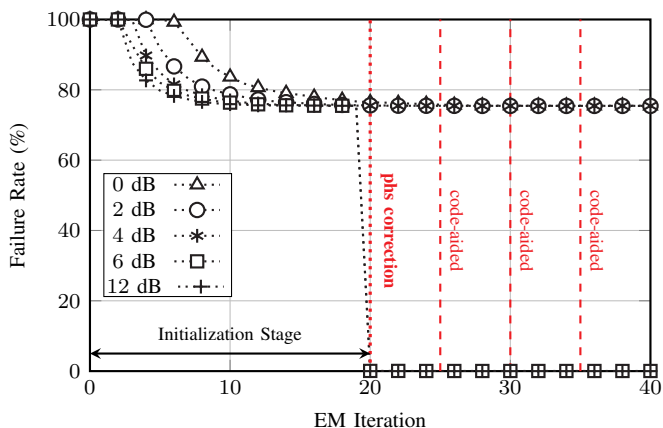} 
    \caption{Failure rate of the phase-aware EM estimator over EM iterations for SNR~$=0$, $2$, $4$, $6$, and $12$~dBs.}
    \label{fig:fail}
    \vspace{-2mm}
\end{figure}

\subsection{Simulation Setup}
All simulation results are obtained from $5000$ independent Monte Carlo runs. A simple rate~$1/2$ convolutional code with generator polynomials $(5,7)_8$ and constraint length $L_c=3$ is applied. A random bit-interleaver of the same length as the codeword is used, and the interleaved bits are subsequently mapped to QPSK symbols. We consider an OFDM specification with a DFT size of $M = 256$, a cyclic prefix size of $N_{cp}=8$, and $N=10$ OFDM symbols per transmission frame. The proposed algorithm is evaluated using frequency-selective channels with time-domain invariant channel coefficients represented as \(\boldsymbol{h} = [0.5, 0.7, 0.5]^\top\). These coefficients are chosen based on their worst-case minimum Euclidean distance at the output, as discussed in \cite[Ch.~9]{Proakis}. To assess our phase-aware EM algorithm, we assume there is an unknown phase \(\theta \sim \mathcal{U}(0, 2\pi)\) that adds to the time-domain channel coefficients, resulting in \(\boldsymbol{h} e^{j\theta}\). We define $\text{SNR}=E\{|X_t|^2\}/\sigma_w^2$, where we assume the noise variance $\sigma^2_w$ is perfectly known. 

For all algorithms, the channel frequency response is first initialized with the 2D estimates $\mathbf{\hat{H}}^{(0)} = {\mathbf{Y}\,\tilde{\mathbf{X}}^{H}} / \lVert \tilde{\mathbf{X}} \rVert^{2}$ where $\tilde{\mathbf{X}}$ is approximated by their hard-decision estimates from the received signal $\mathbf{Y}$ as
\begin{align}
\tilde{X}_{m,n} = 
\begin{cases}
\operatorname{sign}\!\big(\Re\{Y_{m,n}\}\big), 
& \text{if } \big|\Re\{{Y}_{m,n}\}\big | \ge \big|\Im\{{Y}_{m,n}\}\big|,\\
j\,\operatorname{sign}\!\big(\Im\{{Y}_{m,n}\}\big), 
& \text{otherwise.}
\end{cases}\nonumber
\end{align}
During the initialization stage, the EM algorithm is first applied for $20$ EM iterations to produce the common column vector $\boldsymbol{\hat{H}}^{(20)}$ for each OFDM symbol $n$ before entering the turbo decoding module. With the proposed phase-aware EM algorithm, $\boldsymbol{\hat{H}}^{(20)}$ is further refined by $\boldsymbol{\hat{H}}^{(20)}e^{-j\hat{\varphi}}$ to correct any potential phase ambiguities. In each turbo iteration, the EM estimator performs $N_{\text{EM}}=5$ internal EM iterations before passing the estimated likelihoods to the turbo decoding module. Namely, the extrinsic information is incorporated to refine the channel estimates every $5$ EM iterations. We use two metrics for evaluating our system performance: the mean square error (MSE) of the channel response estimates $\text{MSE}=\frac{1}{M}\sum_{m=1}^{M}|{\hat{H}}^{(p)}_m-H_m|^2$ and the failure rate (FR), defined as the percentage of runs where the square error exceeds $10^{-1}$. To ensure the reliability of the detected phase and shift ambiguity, we only perform the initialization refinement if the logarithmic evidence of the selected model $\ln{p(\mathbf{Y}|\circlearrowright i)}$ is at least $10^3$ times larger than the rest of the models.

\subsection{Results and Discussion}
To demonstrate the efficacy of the phase-aware EM algorithm presented in this paper, we used the final estimates $\boldsymbol{\hat{H}}^{(20)}$ from conventional EM, code-aided EM, and phase-aware EM with an SNR of $20$~dB to perform zero-forcing equalization $\mathbf{Y}_{\text{eq}} = \mathbf{Y}/\boldsymbol{\hat{H}}$ and show the constellation diagrams in Figs.~\ref{fig:eq_std}--\ref{fig:eq_phs}. Compared the received data in Fig.~\ref{fig:eq_Y}, all EM-based algorithms can recover the constellation shape. However, conventional and code-aided EM apparently converge to a local maximum where they are unaware of the phase ambiguity. The phase-aware EM, on the other hand, correctly rotates the received symbols to their corresponding constellation order as shown in Fig.~\ref{fig:eq_phs}.

In Figs.~\ref{fig:e05_std}--\ref{fig:e05_joi}, we present the MSE of the channel estimates. As shown in Figs.~\ref{fig:e05_std} and \ref{fig:e05_phs}, the performance exhibits significant variance across iterations due to frequent divergence caused primarily by phase ambiguity. While the code-aided EM improves estimation by incorporating extrinsic information from the decoder (e.g., the MSE reduction after the $25$-th iteration in Fig.~\ref{fig:e05_phs}), realizations that have already converged to a local maximum remain unaffected. By using the phase detection technique, realizations where EM diverges are corrected, resulting in a significant reduction in variance after phase correction at the $20$-th iteration compared to both conventional and code-aided EM approaches. This improvement is evident from the smaller green-shaded area in Fig.~\ref{fig:e05_joi}.

In Fig.~\ref{fig:fail}, the failure rate of our proposed phase-aware code-aided EM estimator is reported with regard to EM iterations. During the initialization stage, the failure rate gradually decreased from $100$\% to $78$\%, indicating that some of the realizations are converging through EM optimization. However, the rate stabilizes after 10 iterations with marginal improvement for SNR $\geq4$~dB. This observation suggests that most realizations converge to a local maximum that the EM algorithm cannot improve further. After incorporating the code-aided phase detection algorithm, the failure rate drops from $78$\% to almost $0$\% when operating at SNR $\geq6$~dB. This demonstrates the robustness of the phase-aware EM for OFDM systems developed in this paper.

\section{Conclusions}\label{sec:conc}
We presented a decoder-aided blind phase ambiguity detection strategy for OFDM systems to address the local maximum problem inherent in the EM algorithm. By incorporating decoder feedback as a model selection criterion and constructing phase candidates according to the PSK modulation format, the algorithm reliably resolves phase ambiguity. The phase-aware EM algorithm developed for OFDM systems operates independently per subcarrier and achieves a significant complexity reduction compared to its single-carrier counterpart. Simulation results confirmed the effectiveness of the algorithm, where the local maximum rate is reduced from nearly $80\%$ in the conventional code-aided EM algorithm to almost $0\%$ when operating with an SNR $\ge6$~dB. 

Future research may extend this framework to dynamic fading environments, where rapid phase fluctuations impose additional challenges on blind initialization, as in practice, each subcarrier could carry an independent phase ambiguity value. Another promising direction is to adapt the proposed scheme to other modulation formats, such as quadrature amplitude modulation (QAM), and to evaluate it with other FEC schemes, thereby broadening its applicability across various mobile communication standards.

\section*{Acknowledgments}
This work was funded by the RAISE collaboration framework between Eindhoven University of Technology and NXP, including a PPS-supplement from the Dutch Ministry of Economic Affairs and Climate Policy.



\end{document}